\documentclass[twocolumn,aps,showpacs,floatfix,amsmath,amssymb,superscriptaddress,prb]{revtex4-2}

\usepackage{graphicx}
\usepackage{bbm}
\usepackage{bm}
\usepackage{dsfont}
\usepackage{mathtools}
\usepackage[colorlinks=true,citecolor=blue,urlcolor=blue]{hyperref}

\newcommand{\bc}{\begin{center}}
\newcommand{\ec}{\end{center}}
\newcommand{\be}{\begin{equation}}
\newcommand{\ee}{\end{equation}}
\newcommand{\ba}{\begin{array}}
\newcommand{\ea}{\end{array}}
\newcommand{\beq}{\begin{eqnarray}}
\newcommand{\eeq}{\end{eqnarray}}
\newcommand{\ket}[1]{\left| {#1}\right\rangle}

\newcommand{\abs}[1]{\left| {#1}\right|}
\newcommand{\trip}[3]{\langle {#1}|{#2}|{#3}\rangle}

\begin{document}

\title{Random singlets and permutation symmetry in the disordered spin-2 Heisenberg chain: A tensor network renormalization group study}
\author{Yen-Tung Lin}
\affiliation{Department of Physics, National Tsing Hua University, Hsinchu 300, Taiwan}

\author{Shao-Fu Liu}
\affiliation{Department of Physics, National Tsing Hua University, Hsinchu 300, Taiwan}

\author{Pochung Chen}
\email{pcchen@phys.nthu.edu.tw}
\affiliation{Department of Physics, National Tsing Hua University, Hsinchu 300, Taiwan}
\affiliation{National Center of Theoretical Sciences, National Taiwan University, Taipei 10617, Taiwan}
\affiliation{Frontier Center for Theory and Computation, National Tsing Hua University, Hsinchu 30013, Taiwan}

\author{Yu-Cheng Lin}
\email{yc.lin@nccu.edu.tw}
\affiliation{Graduate Institute of Applied Physics, National Chengchi University, Taipei 11605, Taiwan}
\affiliation{National Center of Theoretical Sciences, National Taiwan University, Taipei 10617, Taiwan}

\begin{abstract}
We use a tensor network renormalization group method to study random  $S=2$
antiferromagnetic Heisenberg chains with alternating bond strength
distributions.  In the absence of randomness, an imposed dimerization with bond
alternation induces two quantum critical points between the $S=2$ Haldane
phase, a partially dimerized phase and a fully dimerized phase, depending on
the strength of dimerization.  These three phases, called
($\sigma$,$4-\sigma$)=(2,2), (3,1) and (4,0) phases, are valence-bond solid
(VBS) states characterized by $\sigma$ valence bonds (effective spin-1/2
singlet pairs) forming across even links and $4-\sigma$ valence bonds on odd
links.  Here we study the effects of bond randomness on the ground states of
the dimerized spin chain, calculating disorder-averaged twist order parameters
and spin correlations.  We classify the types of random VBS phases depending on
the strength of bond randomness $R$ and dimerization $D$ using the twist order
parameter, which has a negative/positive sign for a VBS phase with odd/even
$\sigma$.  Our results demonstrate the existence of a multicritical point in
the intermediate disorder regime with finite dimerization, where (2,2), (3,1)
and (4,0) phases meet. This multicritical point is at the junction of three
phase boundaries in the $R$-$D$ plane: the (2,2)-(3,1) and (3,1)-(4,0)
boundaries that extend to zero randomness, and the (2,2)-(4,0) phase boundary
that connects another multicritical point in the undimerized limit.  The
undimerized multicritical point separates a gapless Haldane phase and an
infinite-randomness critical line with the diverging dynamic critical exponent
in the large $R$ limit at $D=0$.  Furthermore, we identify the (3,1)-(4,0)
phase boundary as an infinite-randomness critical line even at small $R$, and
find the signature of infinite randomness at the (2,2)-(3,1) phase boundary
only in the vicinity of the multicritical point.
\end{abstract}

\date{\today}

\maketitle

\section{Introduction}
\label{sec:intro}

The ground state properties of antiferromagnetic Heisenberg spin chains have
attracted a lot of attention for many decades, in particular after Haldane’s
conjecture~\cite{Haldane1, Haldane2} that half-integer and integer spin chains
are distinct from each other.  Half-integer spin chains with Heisenberg
interactions have ground-state properties generically similar to the exactly
solvable spin-1/2 chain~\cite{Bethe}, which has a gapless excitation
spectrum~\cite{Pearson} and power-law decaying spin correlations.  Integer spin
chains, on the other hand, have gapped ground states with exponentially
decaying spin correlations~\cite{Haldane1, Haldane2} and a hidden topological
order that is characterized by a nonlocal string correlation
function~\cite{String_1}.  Qualitative differences are also found between
half-integer and integer spin chains  when the coupling constants alternate
between two different values~\cite{Affleck_Haldane,Guo}.  The ground state is
dimerized by infinitesimal alternation for half-integer spins, while the
Haldane phase in an integer spin chain changes to a dimer state via a phase
transition only with sufficiently strong bond alternation.

Another theme of high interest in condensed matter physics is the interplay
between disorder, interactions and quantum fluctuations. The ground-state
properties of low-dimensional quantum systems are often modified dramatically
by introducing quenched disorder (i.e. time-independent disorder). Remarkably,
there are properties resulting from quenched disorder that are universal for
a broad class of quantum spin chains, independently of whether the spin is
integer or half-integer.  The so-called random-singlet (RS)
phase~\cite{Fisher_AF} is one of the disorder induced phases that possesses such
universal properties. This phase describes the ground state of a spin-1/2
Heisenberg chain with any amount of disorder in couplings, and is also the
ground-state phase of the spin-1 chain in the strong disorder limit where the
excitation spectrum becomes gapless and the string topological order
vanishes~\cite{Hyman,Monthus_PRL,Monthus_PRB,Lajko,Tsai}.

The RS phase was first found in the ground state of the random
spin-1/2 chain by using the strong-disorder renormalization group (SDRG)
method~\cite{Ma,Dasgupta,Fisher_AF,SDRG_rev}. The SDRG method for the spin-1/2 chain
consists of iteratively
locking the strongest coupled spin pair into a singlet state, which decouples
from the rest of the chain after effective couplings are generated among the
remaining spins.  This SDRG scheme ultimately flows toward an 
RS fixed point~\cite{Fisher_AF} that asymptotically represents the system's ground state,
in which
pairs of strongly entangled spins form singlets over all length scales, 
mostly short ranged but occasionally very long ranged.
Those long-ranged singlets are rare; however, they dominate the {\it average}
spin-spin correlations that decay asymptotically with distance $L$ as an inverse-square form $L^{-2}$.
By contrast, the correlations between typical pairs of widely separated spins  
are very weak and
decay exponentially with the square root of their distance.
Furthermore, 
the characteristic energy scale, $\epsilon$, and length scale, $L$, of the singlets
in the RS phase follow: 
\be
     -\ln \epsilon \sim L^\psi\,,
     \label{eq:lnE_L}
\ee  
with $\psi=1/2$. This energy-length scaling is very different from the standard dynamic scaling,
$\epsilon \sim L^{-z}$, and implies that the dynamic exponent diverges $z\to\infty$. 
The RS fixed point is one example of an infinite-randomness fixed point~\cite{Motrunich},
which is characterized by extremely broad distributions of physical properties, even on a logarithmic scale,
leading to the distinction between average and typical behaviors.

\begin{figure}[tb!]
\centerline{\includegraphics[width=8.6cm]{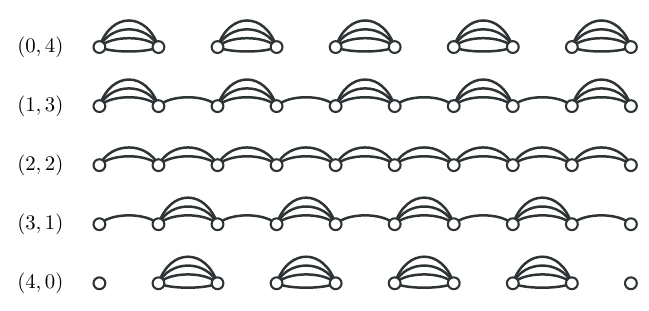}}
\caption{Five distinct VBS phases ($\sigma$, 4-$\sigma$) with $\sigma=0,1,2,3,4$
for a spin-2 chain, where $\sigma$ denotes the number of spin-1/2 singlets (indicated by the arches)
over each even bond.
}
\label{fig:vbs}
\end{figure}

The SDRG method has been extended to higher spin
chains~\cite{Hyman,Monthus_PRL,Monthus_PRB,Damle,Damle_Huse,S23}. In particular, Damle and Huse
applied an extended SDRG scheme to disordered spin-$S$ Heisenberg chains with arbitrary
$S$ and obtained a class of infinite-randomness fixed points, called permutation symmetric fixed points~\cite{Damle_Huse}. 
In the valence-bond (VB) picture for spin $S > 1/2$, each spin $S$ is replaced with $2S$ virtual 
spin-1/2 variables, and valence bonds (singlets) are pairwise created between spin-1/2 variables that belong 
to different spin-$S$ sites. 
For a spin-$S$ chain, there are $2S+1$ distinct valence-bond solid (VBS) domains,
denoted by $(\sigma, 2S-\sigma)$ with $\sigma\in\{0,1,\cdots,2S\}$;
a VBS domain of type $(\sigma, 2S-\sigma)$ (type $\sigma$)  consists of  $\sigma$  spin-1/2 singlets
over each even bond and $2S-\sigma$ spin-1/2 singlets over each odd bond (Fig~\ref{fig:vbs}).
Among these generalized VBS states, the Haldane phase in an integer spin-$S$ chain is associated with
the symmetric $(S, S)$ domain. 
A dimerized state $(\sigma, 2S-\sigma)$ with $\sigma \neq S$ 
can be realized 
in chains with alternating bond strength distributions.
In the extended SDRG scheme, the degrees of freedom 
are effective spins of magnitude 
\[
    S_{\sigma,\sigma'}=\frac{\abs{\sigma-\sigma'}}{2}\,,
    \label{eq:dw_spin}
\]
localized at the boundaries between distinct domains of type $\sigma$ and $\sigma'$.
These domain-wall spins interact with spins in neighboring domain walls through effective couplings
that can be antiferromagnetic or ferromagnetic, depending on the types of environment domains.   
The renormalization of strong effective couplings between domain-wall spins under the RG leads to 
reconfigurations of associated domains.
Using this domain-wall picture, Damle and Huse have predicted a series of infinite-randomness multicritical points,
$\mathcal{P}_n$, that result from a competition between $n$ domains. 
At a $\mathcal{P}_n$ multicritical point, $n$ domains appear with equal probability; hence the
multicritical point is called permutation symmetric critical point. 
For a $\mathcal{P}_n$ multicritical point, we have  
\be
     \psi_n=1/n\,,
     \label{eq:psi}
\ee
for the energy-length scaling exponent given in Eq.~(\ref{eq:lnE_L});
this is a generalization of
the RS fixed point with $n=2$, where two domains occur with equal probability under the action of the RG.

The Griffiths singularity (also known as the Griffiths-McCoy singularity for quantum systems)~\cite{Griffiths,McCoy,McCoy_Wu} 
is another interesting phenomenon arising from
the interplay between quantum fluctuations and randomness. This phenomenon is characterized by
singular low-energy behavior of various observable, such as the susceptibility and the specific heat, 
in an off-critical phase.
Griffiths effects occur 
when there are
rare but arbitrarily large spatial regions that are locally in a phase $B$ due to 
disorder fluctuations, while the system is overall in a phase $A$.
Quantum fluctuations enhance the effects of the rare regions, leading to
a power-law behavior of the density of low-energy excitations such that~\cite{Vojta_rev}
\be
     \rho(\epsilon)\sim \epsilon^{-1+1/z} 
     \label{eq:dos}
\ee
for a one-dimensional system, with a non-universal continuously varying exponent $z$
that also describes the length scale and the energy scale through $\epsilon \sim L^{-z}$.
This power-law density of states is responsible for power-law singularities of certain observables at low temperatures.
Using the SDRG analysis~\cite{Hyman,Hyman_Dimer,Damle}, the generalized VBS states in dimerized spin-$S$ chains, 
including dimer phases for any $S$ and the Haldane phase in an
integer-$S$ chain, in the presence of sufficiently strong disorder have been identified as
Griffiths phases, 
where the energy gaps are filled in yet the topological order (Haldane order and dimer order) persists.

Considerable numerical efforts have been devoted to examine the theoretical
predictions about the ground-state phases of random spin $S>1/2$ chains, but so
far have been mainly focused on the $S=1$
chain~\cite{QMC_spin1,Lajko,Torlai,Tsai,Twist_spin1}.  Here we first use a
tensor network strong-disorder renormalization group (tSDRG) method~\cite{tSDRG1} to study
the zero-temperature phases of the random spin-2 chain with alternating bond
strength distributions, which to our knowledge have not been previously
investigated numerically.

The paper is organized  as follows. In Sec.~\ref{sec:model} we define the model
and summarize some known properties of the ground-state phases. In
Sec.~\ref{sec:method} we outline the SDRG and tSDRG methods.  In
Sec.~\ref{sec:results} we provide our numerical results for the ground-state
phases depending on randomness and dimerization, focusing on a VBS order
parameter and end-to-end correlations.  We conclude in Sec.~\ref{sec:summary}
with a summary and discussion.

\section{The model}
\label{sec:model}

\begin{figure*}[tbh!]
\includegraphics[width=16cm, clip]{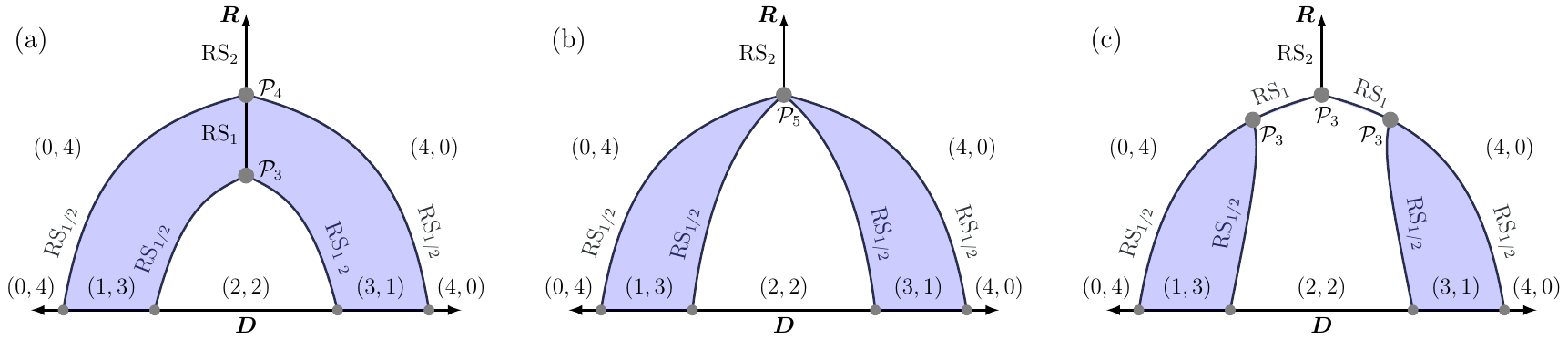}
\caption{\label{fig:diagram}
Three possible phase diagrams of spin-2 chains in the $R$-$D$ plane, predicted in Ref.~\cite{Damle_Huse}. 
}
\end{figure*}

The model we consider is the spin-2 antiferromagnetic Heisenberg chain, described
by the Hamiltonian:
\be
   H =  \sum_{i} J_{i}\, \mathbf{S}_{i}\cdot\mathbf{S}_{i+1}\,,
   \label{eq:H}
\ee
where $\mathbf{S}_{i}$ is the spin-2 operator at site $i$,
and $J_i$ are random coupling constants given by
\be
    J_i=K_i[(1+(-1)^i D]\,,
    \label{eq:J}
\ee
where the parameter $D$, with $\abs{D}\le 1$, measures the strength of bond alternation (dimerization),
and  $K_i$ are random positive variables with 
the following  power-law  distribution:
\be
    P(K) = \begin{cases} 
    R^{-1} K^{-1+1/R},\quad \text{for }\, 0\le K \le 1\,,\\
    0, \quad \text{otherwise}. 
    \end{cases}
     \label{eq:PJ}
\ee
where $R>0$, being the standard deviation of $\ln(K)$, 
parameterizes the strength of the randomness.
This power-law distribution of bond randomness at $D=0$
has been widely used in previous numerical studies for
disordered systems~\cite{2DAF,Lajko,Laflorencie1D,Twist_spin1,Tsai}.

The spin-2 model hosts a variety of ground-state phases, depending upon the
dimerization $D$ and the randomness $R$.  In the absence of randomness (i.e.
$R=0$), there are five gapped states; they are the VBS phases of type
$\sigma=0,1,2,3,4$, changing successively as the dimerization parameter $D$
varies from $-1$ to $1$.  The valence bond picture suggests that the elementary
excitations at all four domain walls are effective spin-1/2 variables and the
phase transitions belong to the level-1 SU(2) Wess-Zumino-Witten universality
class~\cite{Affleck_Haldane,WZW2}.  The locations of the domain walls have been determined
using the level-crossing method~\cite{Crossing} and the ground-state
expectation value of a so-called twist operator~\cite{Twist}. The critical
values for the $(2,2)-(3,1)$ and $(3,1)-(4,0)$ transitions are found at
$D_{c,1}\approx 0.18$ and $D_{c,2}\approx 0.55$, respectively.  Under
interchanging even and odd bonds, the $(1,3)-(2,2)$ and $(0,4)-(1,3)$
transitions occur at $-D_{c,1}$ and $-D_{c,2}$, respectively.

According to the Damle-Huse domain-wall picture~\cite{Damle_Huse},
random-singlet RS$_S$ phases out of spin $S=1/2, 1, 2$ variables and
multicritical points $\mathcal{P}_n$ with possible $n=3,4,5$ can occur when
disorder comes into play.  Since the undimerized spin-1/2 chain with any weak
randomness is in the RS$_{1/2}$ phase, the critical domain walls at
$\abs{D_{c,1}}$ and $\abs{D_{c,2}}$ with effective spin-1/2 are expected to
evolve into an RS$_{1/2}$ state for $R>0$.  On the other hand, the RS$_2$
phase, arising from a competition between the $(4,0)$ domain and the $(0,4)$
domain, occurs only in the strong disorder limit at $D=0$.  In the  RS$_2$
phase, spin-2 singlets connect sites on different sublattices over arbitrarily
long distances in a random fashion, completely analogous to the RS$_{1/2}$
state.

For the spin-2 chain, there are three possible arrangements of random-singlet RS$_S$ phases and
multicritical points $\mathcal{P}_n$ in the $R$-$D$ plane between weak and very strong disorder, as shown in Fig.~\ref{fig:diagram},
among which the phase diagram with the maximally symmetric multicritical point $\mathcal{P}_5$ may occur 
only with some additional fine-tuned parameter in the model~\cite{Damle_Huse}, {or can be realized only 
with higher symmetry groups~\cite{Hoyos1,Hoyos2,Hoyos3}.} 
All multicritical points $\mathcal{P}_n$ belong to a set of infinite-randomness fixed points
with the critical exponent $\psi_n$
given in Eq.~(\ref{eq:psi}), and the correlation-length exponent given by~\cite{Hyman,Monthus_PRL,Damle_Huse}
\be
      \nu_n=\frac{2}{\sqrt{4n+1}-1} \frac{1}{\psi_n}\,.
      \label{eq:nu}
\ee
In particular, the random-singlet RS$_S$ phases for all $S$
is the special case with $n=2$, where we have $\psi=1/2$ and $\nu=2$. 

The off-critical regions, including the Haldane phase associated with the $(2,2)$ domain and the dimerized phases,
have excitation gaps and exhibit topological order in the absence of randomness.
The topological order in the Haldane phase can be detected by a nonzero limiting value of the generalized string order parameter~\cite{String_1,String_S}
\begin{equation}
       O^{z}_{j,k} = -\left \langle S_{j}^{z} \exp(i \theta \sum_{l = j + 1}^{k - 1} S_{l}^{z} ) S_{k}^{z} \right\rangle \, 
        \label{eq:Oz}
\end{equation}
in the $\abs{j-k}\to\infty$ limit, 
where $S^z_j$ is the $z$ component of the spin-2 operator at site $j$, and an angle parameter with $\theta=\pi/2$ 
is mostly suitable for the spin-2 case
since $O^{z}_{j,k}$ becomes a smooth function of the distance $\abs{j-k}$~\cite{String_2},
similar to $\theta=\pi$ for the spin-1 case~\cite{String_1}.
The topological order in the dimer phases is enforced by the Hamiltonian, and its sign reflects whether
the Hamiltonian favors singlet pairs to be formed over even or odd bonds.
Both the Haldane topological order and dimer order can survive in the presence of randomness even when the energy gaps close up .
Such a gapless region with topological order is the Griffiths phase.  

The main goal of our numerical study is to identify which diagram in Fig.~\ref{fig:diagram} corresponds to the phase diagram 
of our spin-2 random chain.

\section{The numerical method}
\label{sec:method}

\subsection{SDRG}

The  Hamiltonian of a disordered quantum many-body system has an intrinsic
separation in energy scales that allows us to find its ground state using the
SDRG technique.  The essential idea of the SDRG method, introduced in
Ref.~\onlinecite{Ma,Dasgupta}, is to find the largest term in the Hamiltonian successively
and put the subsystem associated with this term into its ground state.
The couplings between this subsystem and the rest of the system 
are treated by perturbation theory and effective couplings across the subsystem are generated.
For example, for the random spin-1/2 Heisenberg antiferromagnetic chain, an effective coupling
is generated between spins $k-1$ and $k+2$ with strength 
\be
     \tilde{J} = \frac{1}{2} \frac{J_{k-1}J_{k+1}}{J_{k}}\,,
     \label{eq:Jeff}
\ee  
when $J_k$ is the strongest coupling that locks spins $k$ and $k+1$ into a singlet at a certain step of RG. 
The new energy scale is then the strength of the strongest remaining coupling: $\Omega=\max\{\tilde{J}\}$.
By repeating this RG procedure, we gradually lower the energy scale
and
reduce the number of degrees of freedom in the system.
The RG flow equation describing the evolution of the probability distribution under the RG process 
has been solved by D.~S.~Fisher in Ref.~\cite{Fisher_AF} for the spin-1/2 chain.
The multiplicative relation in Eq.~(\ref{eq:Jeff}) suggests that
it is more convenient to measure bond strengths on a logarithmic scale.
In terms of logarithmic variables defined as $\Gamma=-\log(\Omega)$ and $\zeta=\log(\Omega/\tilde{J})$,
the fixed point distribution for the spin-1/2 chain corresponds to 
\be
           P(\zeta)=\frac{1}{\Gamma}\,e^{-\zeta/\Gamma}\,,
\ee

For higher spin chains,
renormalized couplings in the conventional SDRG method
may become stronger than the decimated couplings when the randomness is not
sufficiently strong, which makes perturbation theory invalid.
Thus, several extended SDRG methods, based on effective $S=1/2$ models with both antiferromagnetic and ferromagnetic couplings, 
have been proposed for higher spin chains~\cite{Hyman,Monthus_PRL,Monthus_PRB,Refale_S1,Damle_Huse}.  
Using a domain-wall model, Damle and Huse have extended Fisher's RG analysis to arbitrarily high spin.
In the domain-wall model, one defines $\rho_\sigma$ to be the probability that a specific domain is of type $\sigma$
and $W_{\sigma \sigma'}$ to be the probability of a domain of type $\sigma$ followed by one of type $\sigma'$,
which are required to formulate the RG flow equations.
The fixed point solution, which controls the multicritical point $\mathcal{P}_n$, is found to be given by 
the bond strength distribution:
\be
    P_\sigma(\zeta) =\frac{n-1}{\Gamma}\,e^{-(n-1)\zeta/\Gamma}\,,\quad\forall \sigma
\ee 
and
\be
 \begin{split}
    &\rho_\sigma =1/n,\quad \forall \sigma\,, \\
    &W_{\sigma \sigma'} = 1/(n-1),\quad  \forall \sigma\neq \sigma'\,,
 \end{split} 
\ee
indicating the domain permutation symmetry. 
From the fixed point solution, one can deduce the energy-length scaling relation $-\ln \epsilon \sim L^{\psi_n}$
with $\psi_n$ given in Eq.~(\ref{eq:psi}).

\subsection{Tensor networks and SDRG}
Here we use a tree tensor-network generalization of the SDRG, referred to as tSDRG,
to study the spin-2 random chain.
This generalized SDRG scheme, proposed by Goldsborough and R\"omer~\cite{tSDRG1}, formulates the RG procedure as a tree tensor network
and refines the perturbative approximation of SDRG
by including higher energy states at each RG step,
along the lines of a previous SDRG extension~\cite{Hikihara}.
The tSDRG method has been applied in studies of the quantum Ising chain~\cite{tSDRG2}
and spin-1 chains~\cite{Tsai}, where accurate results that are compatible with
the results obtained by non-approximate quantum Monte Carlo calculations~\cite{Shu} and
the density matrix renormalization group~\cite{Lajko,Torlai} are achieved.

The starting point of tSDRG is to express a one-dimensional Hamiltonian of $L$ sites as a sum of two-site Hamiltonian
in terms of matrix product operators (MPOs)~\cite{Schollwock}:
\begin{equation}
  H = \sum_i H_{i,i+1} = W^{[1]} W^{[2]} \cdots W^{[L]}\,,
\end{equation}
where an MPO $W^{[i]}$ at site $i$ is a matrix of operators.
Specifically for the model considered in this work, the two-site Hamiltonian reads
\begin{equation}
\begin{split}
  H_{i,i+1} &= J_i \mathbf{S}_{i}\cdot\mathbf{S}_{i+1} \\
            &= J_{i}[\frac{1}{2}({S}_{i}^{+} {S}_{i+1}^{-} + {S}_{i}^{-} {S}_{i+1}^{+} ) + {S}_{i}^{z} {S}_{i+1}^{z}]\,.
\end{split}
\end{equation}
where $S^{+}$ and $S^{-}$ are the ladder operators. For a chain with open boundary conditions (OBC),
we have the following explicit form of the $W$ tensors:
\begin{equation}
\label{eq:MPO2}
W^{[i]} =
\begin{pmatrix}
        {1} & 0 & 0 & 0 & 0 \\
        {S}_{i}^{+} & 0 & 0 & 0 & 0 \\
        {S}_{i}^{-} & 0 & 0 & 0 & 0 \\
        {S}_{i}^{z} & 0 & 0 & 0 & 0 \\
        0 & (J_{i}/2){S}_{i}^{-} & (J_{i}/2){S}_{i}^{+} & J_{i}{S}_{i}^{z} & {1} \\
\end{pmatrix}\ ,
\end{equation}   
for a site in the bulk, i.e. $i \neq 1, L$, and
\begin{equation}
\label{eq:MPO1}
W^{[1]} =
\begin{pmatrix}
        0\; & (J_{1}/2){S}_{1}^{-}\; & (J_{1}/2){S}_{1}^{+}\; & J_{1}{S}_{1}^{z}\; & {1} \\
\end{pmatrix}\ ,
\end{equation}

\begin{equation}
\label{eq:MPO3}
W^{[L]} =
\begin{pmatrix}
        {1}  \\
        {S}_{L}^{+}  \\
        {S}_{L}^{-} \\
        {S}_{L}^{z} \\
        0  \\
\end{pmatrix}\ .
\end{equation}
for two edge sites $i=1$ and $L$.
For a chain with periodic boundary conditions (PBC), the MPO tensors for $i=1\cdots
L$ are all bulk tensors as given in Eq.~(\ref{eq:MPO2}), where the coupling
$J_L$ links between two end sites $L$ and $1$.
An important observation is that the two-site Hamiltonian $H_{i,i+1}$ is encoded in the local matrix product $W^{[i]} W^{[i+1]}$.
It is easy to verify that
\begin{equation}
  W^{[i]} W^{[i+1]} =
  \left( \begin{array}{ccccc}
    1 &  0 & 0 & 0 & 0 \\
    S^+_i &  0 & 0 & 0 & 0 \\
    S^-_i &  0 & 0 & 0 & 0 \\
    S^z_i &  0 & 0 & 0 & 0 \\
    H_{i,i+1} &  \frac{J_{i+1}}{2} S^-_{i+1} & \frac{J_{i+1}}{2} S^+_{i+1} & J_{i+1} S^z_{i+1} & 1
  \end{array} \right).
\end{equation}
That is $H_{i,i+1} = \left( W^{[i]} W^{[i+1]} \right)_{5,1}$.
The essential information required for the tSDRG procedure is the list of MPO $W^{[i]}$ and the list of two-site Hamiltonian $H_{i,i+1}$.

\begin{figure}[tbp!]
\centerline{\includegraphics[width=8cm]{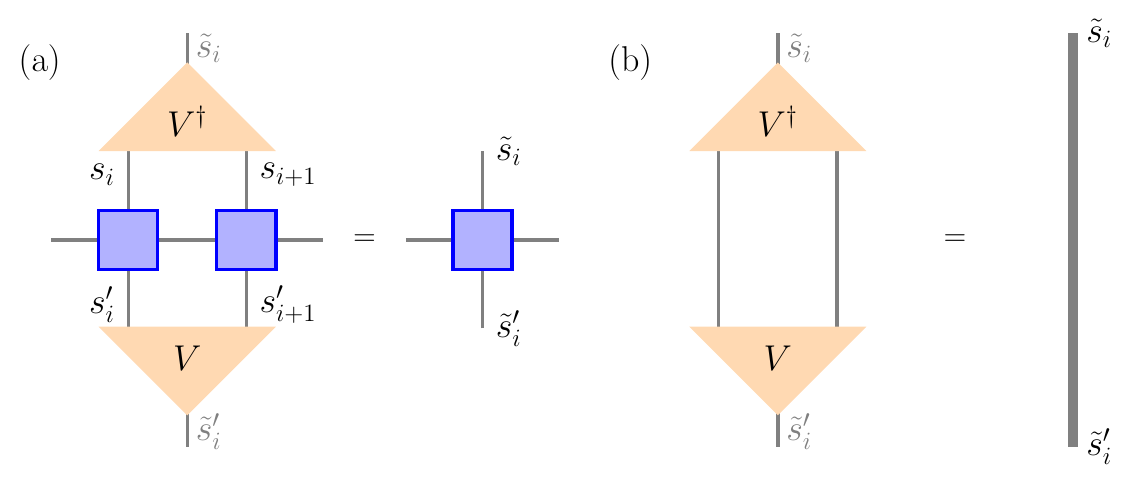}}
\caption{\label{fig:V}
(a) The three-leg tensor $V$ (triangles), built from the $\chi'$ lowest energy eigenvectors,
truncates a two-site tensor into a renormalized site. The blue-shaded squares are
local tensors of the MPO. The vertical legs denote physical indices;
the horizontal legs denote virtual indices; (b) Isometric property $V^\dag V=1$ of the tensor $V$.
}
\end{figure}

Similar to the conventional SDRG, in each RG iteration 
one selects a pair of adjacent sites to be renormalized, depending on the local energy spectrum.
Here the selection is based on the largest energy gap, rather than the strongest coupling (corresponding to the lowest gap)~\cite{Westerberg_PRL, Westerberg_PRB,
Hikihara}.
For each two-site Hamiltonian $H_{i,i+1}$ we identify the energy gap $\Delta_{i,i+1}$, 
which is measured as the difference between the highest energy of the 
$\chi^\prime$-lowest eigenstates that would be kept and the energy
of the $(\chi'+1)$-th eigenstate. 
We set a {\em bond dimension} $\chi$  as the upper bound of the number $\chi^\prime$  to control the accuracy
of the calculation. The actual number $\chi^\prime$ is adjusted to keep full SU(2) multiplets.
By increasing the bond dimension $\chi$, we can obtain more accurate results but at the cost of computational resources and time.

\begin{figure*}[tbp!]
\includegraphics[width=10cm, clip]{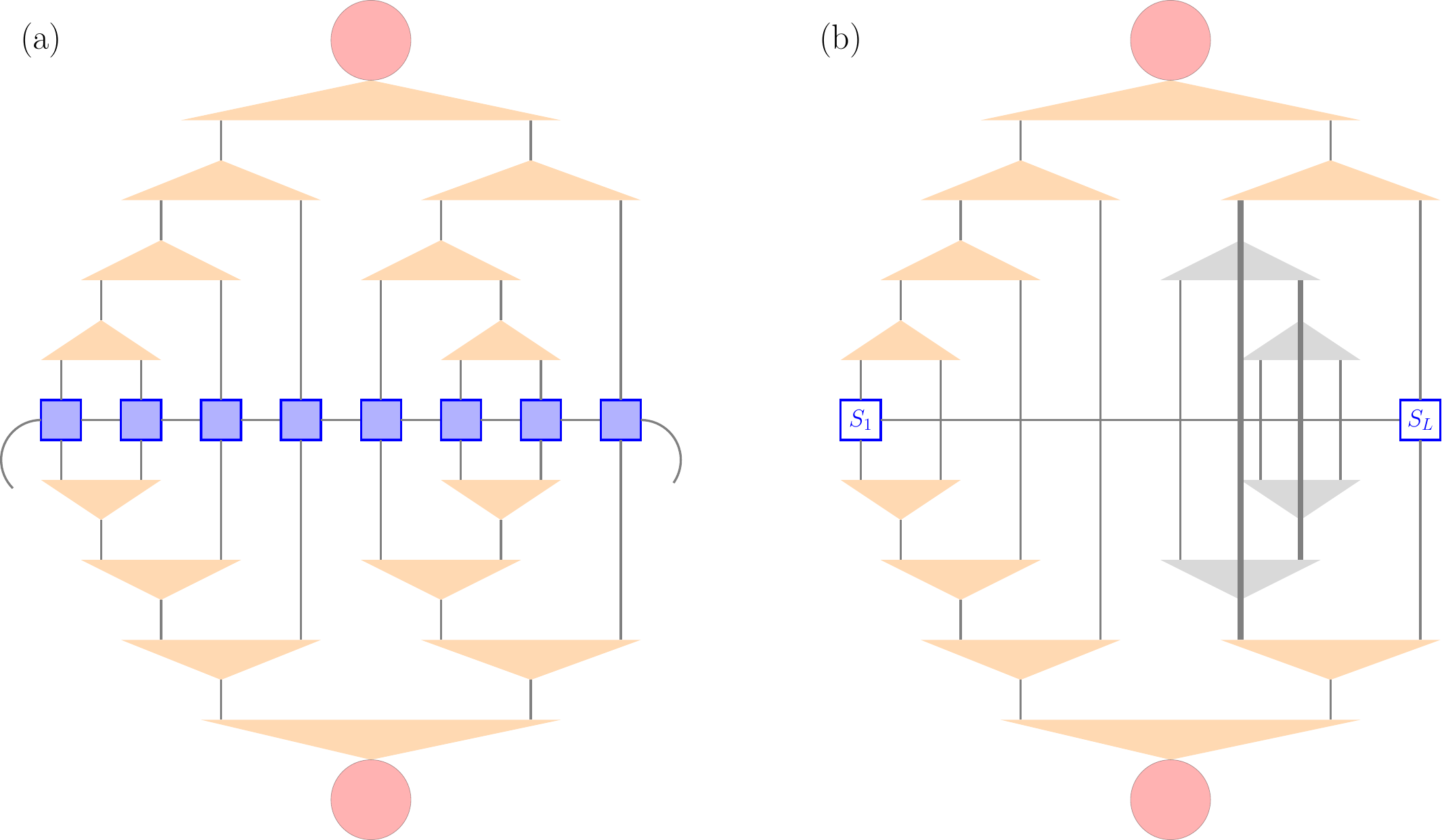}
\caption{\label{fig:tsdrg}
(a) The tSDRG algorithm as a tree tensor network for a chain of 8 sites with periodic boundary conditions. The squares indicate the 
$W$-tensors in the MPO representation of the system's Hamiltonian,
the triangles are the $V$-tensors used to truncate the two-site operators, and the circles represent  
the top Hamiltonian encoded in the final two-site tensor. The RG iteration proceeds upwards in the vertical dimension. 
The part below the W-tensors is the conjugate of the upper part. 
(b) A tree tensor network for the ground-state expectation value of the end-to-end correlator which acts only on two edge sites.
Since $V^\dag V=1$, only those isometric tensors (orange triangles) that affect the two edge sites are considered.
}
\end{figure*}

After obtaining all local gaps described above we identify the two sites with largest gap.
These two sites will be merged into a renormalized site as follows.
We first express the $k$-th eigenstate $|\Psi_k\rangle$ of $H_{i,i+1}$ in terms of the local two-site product basis
\begin{equation}
  |\Psi_k\rangle = \sum_{s_i, s_{i+1}} \Psi_{k,s_i s_{i+1}} |s_i\rangle |s_{i+1}\rangle,
\end{equation}
where $|s_i\rangle |s_{i+1}\rangle$ is the local product basis.
We then construct a projector that projects the two-site space to the renormalized site space with dimension $\chi^\prime$,
\begin{equation}
  V \equiv \sum_{k=1}^{\chi^\prime} |s_i\rangle |s_{i+1}\rangle \Psi_{k, s_i s_{i+1}} \langle \Psi_k|.
\end{equation}
In terms of  matrix notation one can write $V$ as
\begin{equation}
  V = \left( \begin{array}{cccc}
    |\Psi_1\rangle & |\Psi_2\rangle & \cdots  & |\Psi_{\chi^\prime}\rangle
    \end{array} \right)\,,
\end{equation}
and its tensor network diagram is given in Fig.~\ref{fig:V}.
Note that the tensor $V$ has the isometric property $V^\dagger V=1$.
To obtain the renormalized MPO $\tilde{W}^{[i,i+1]}$ associated with the renormalized site,
one uses $V$ to renormalize each element of the product $W^{[i]} W^{[i+1]}$ as follows
\begin{equation}
  \left( W^{[i]} W^{[i+1]} \right)_{b,b'} \rightarrow \tilde{W}^{[i,i+1]}_{b,b'} \equiv V^\dagger \left( W^{[i]} W^{[i+1]} \right)_{b,b'} V. 
\end{equation}
The two-site Hamiltonians that contain the renormalized site can be decoded as follows:
\begin{equation}
  I_{i-1} \otimes H_{i,i+1} \rightarrow \tilde{H}_{i-1, [i,i+1]} = \left( W^{[i-1]} \tilde{W}^{[i,i+1]} \right)_{5,1},
\end{equation}
\begin{equation}
  H_{i,i+1} \otimes I_{i+2}  \rightarrow \tilde{H}_{[i,i+1], i+2} = \left( \tilde{W}^{[i,i+1]} W^{[i+2]} \right)_{5,1}.
\end{equation}
This completes one iteration of tSDRG.
We have now an updated list of MPOs and two-site Hamiltonians. 
Conceptually, one can re-label the sites so that the renormalized MPO and two-site Hamiltonian
are labeled as $W^{[i]}$ and $H_{i,i+1}$ respectively.
In practice the relabeling is not necessary.

To obtain the ground state of the system, we should repeat the tSDRG iteration
until the system contains only two renormalized sites.  At this stage we
diagonalize the top Hamiltonian $H_{\text{top}}$ to obtain its ground state
$|\Psi^{\text{top}}_1\rangle$.  From $|\Psi^{\text{top}}_1\rangle$ and the
projector $V$ at each iteration, one can generate an inhomogeneous binary tree
tensor network as sketched in Fig.~\ref{fig:tsdrg}(a).  The expectation value
of a product of local operators can be obtained by contracting the operators with
the tree and its conjugate as sketched in Fig.~\ref{fig:tsdrg}(b).  The
contraction can be evaluated efficiently thanks to the property that $V^\dagger
V=1$.

\section{Numerical results}
\label{sec:results}

\begin{figure}[tbp!]
\includegraphics[width=10cm, clip]{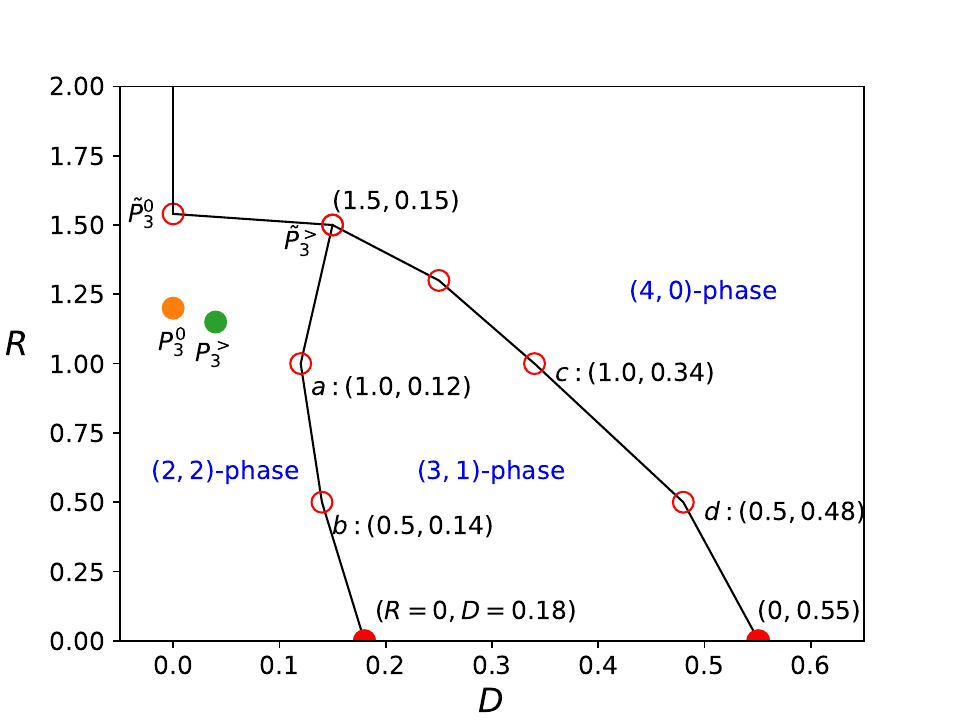}
\caption{\label{fig:z0}
{The phase diagram in the $R$-$D$ plane with $D\ge 0$, extracted from our numerical results (unfilled circles) for $L=512$ and
previous numerical data~\cite{Twist} at $R=0$. The points labeled by $\tilde{P}_3^0$ and $\tilde{P}_3^>$ are 
three-phase multicritical points for $L=512$ at $D=0$ and $D>0$, respectively. 
The points labeled by ${P}_3^0$ and ${P}_3^>$ 
are the multicritical points estimated for $L\to\infty$. 
The four points denoted by $a, b, c$ and $d$
are finite-size critical points whose properties are further discussed in Sec.~\ref{sec:twist} and Sec.~\ref{sec:cor}.} 
}
\end{figure}

In this section we use the tSDRG method to explore ground-state phases of the random spin-2
chain with alternating bond strength distributions as defined in
Eqs.~(\ref{eq:J}) and (\ref{eq:PJ}),  
focusing on two observables: the VBS order parameter based on a unitary operator
appearing in the Lieb-Schultz-Mattis theorem~\cite{LSM} and the end-to-end correlation function.
{We first show the $R$-$D$ phase diagram, extracted from our numerical data, in Fig.~\ref{fig:z0}
before presenting detailed results. We restrict ourselves to the $D\ge 0$ region because the results for $D<0$ can be obtained simply
via the parity symmetry. The phase diagram that we have identified largely agrees with the type in Fig.~\ref{fig:diagram}(c).}

\subsection{VBS order parameter}
\label{sec:twist}

We consider a unitary operator, called the twist operator, defined for a chain with $L$ spins as
\be
   U=\exp\left[i\frac{2\pi}{L}\sum_{j=1}^L j S_j^z \right]\,,
   \label{eq:U}
\ee
which creates spin-wave-like excitations by rotating each spin about the $z$ axis with
a relative angle.
The twist operator was first introduced in the Lieb-Schultz-Mattis theorem~\cite{LSM,Affleck_Lieb,Oshikawa_Affleck}, 
which states that for the ground state, $\Psi_{\text{GS}}$, of a half-integer spin chain, 
one has
\be
     z_L\equiv \trip{\Psi_{\text{GS}}}{U}{\Psi_{\text{GS}}}=0\,,
\ee 
in the limit $L\to\infty$, indicating a gapless excitation spectrum.
Furthermore, the ground-state expectation value of this operator  
has been found to be
capable of detecting and characterizing VBS order~\cite{Twist}.
For a VBS state of type-$\sigma$, the asymptotic form of the expectation value 
is given by~\cite{Twist}
\be
     z_L = (-1)^\sigma \left[1-\mathcal{O}(1/L) \right]\,,
\ee
that is, it
is positive (negative) in the $L\to\infty$ limit
if $\sigma$ is even (odd).
Using the properties of $z_L$, ground-state phase diagrams of dimerized spin-$S$ chains
with $S=1/2,1,3/2$, and $2$ in the absence of randomness were determined in Ref.~\cite{Twist}.
Remarkably, this order parameter is applicable also for strongly disordered systems,
as demonstrated in a quantum Monte Carlo study for the random spin-1 Heisenberg chain~\cite{Twist_spin1}.

\begin{figure}[tb!]
\centerline{\includegraphics[width=8.6cm]{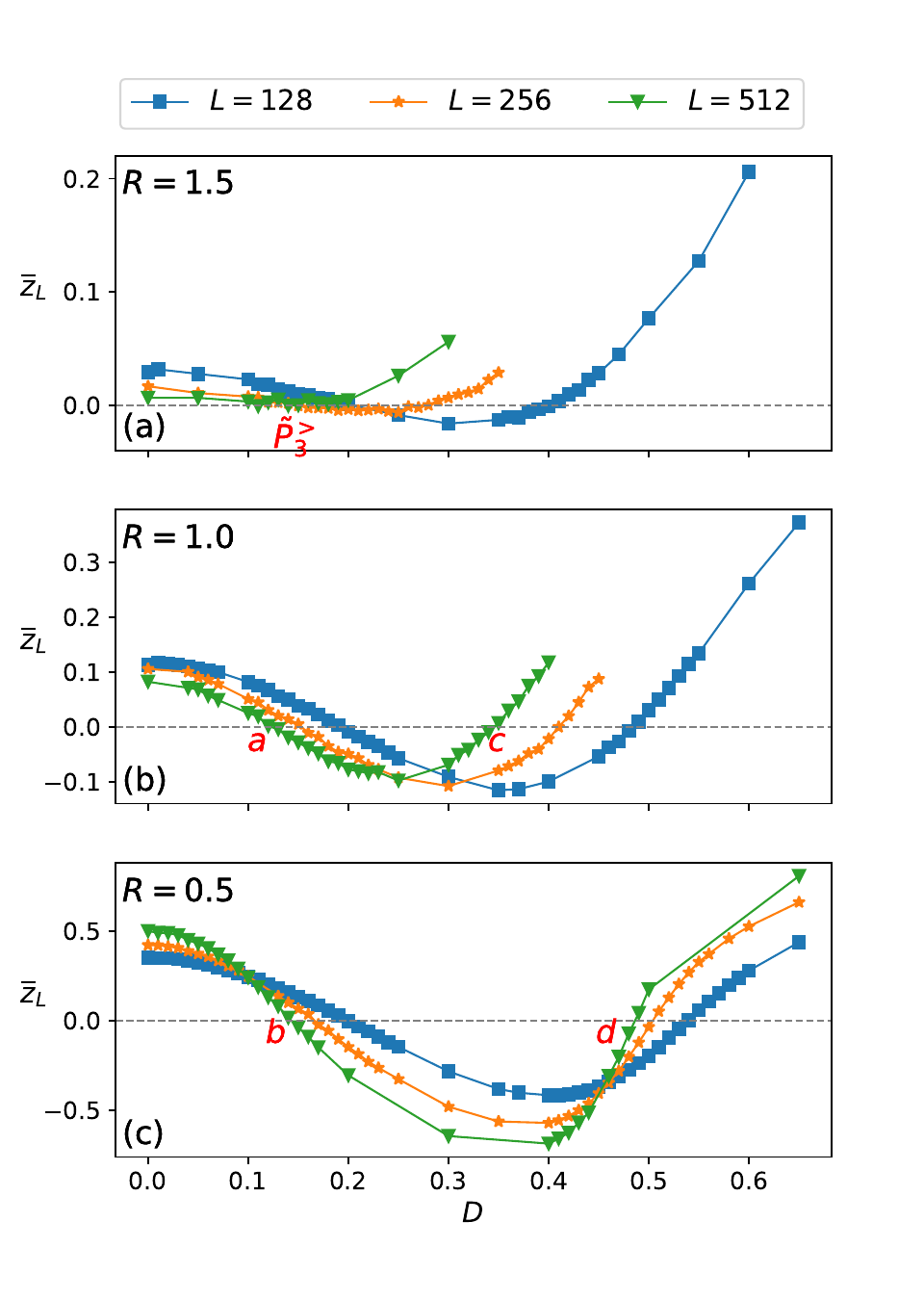}}
\caption{Disorder-averaged twist order parameters $\overline{z}_L$ for different system sizes
at $R=1.5$, $R=1.0$ and $R=0.5$, plotted versus the dimerization $D$. The red labels $\tilde{P}_3^>, a, b, c$
and $d$ indicate the points where $\overline{z}_L=0$ for $L=512$ (green data); these points are also shown
in Fig.~\ref{fig:z0}. 
}
\label{fig:z_D}
\vskip-3mm
\end{figure}

\begin{figure}[tbp!]
\centerline{\includegraphics[width=8.6cm]{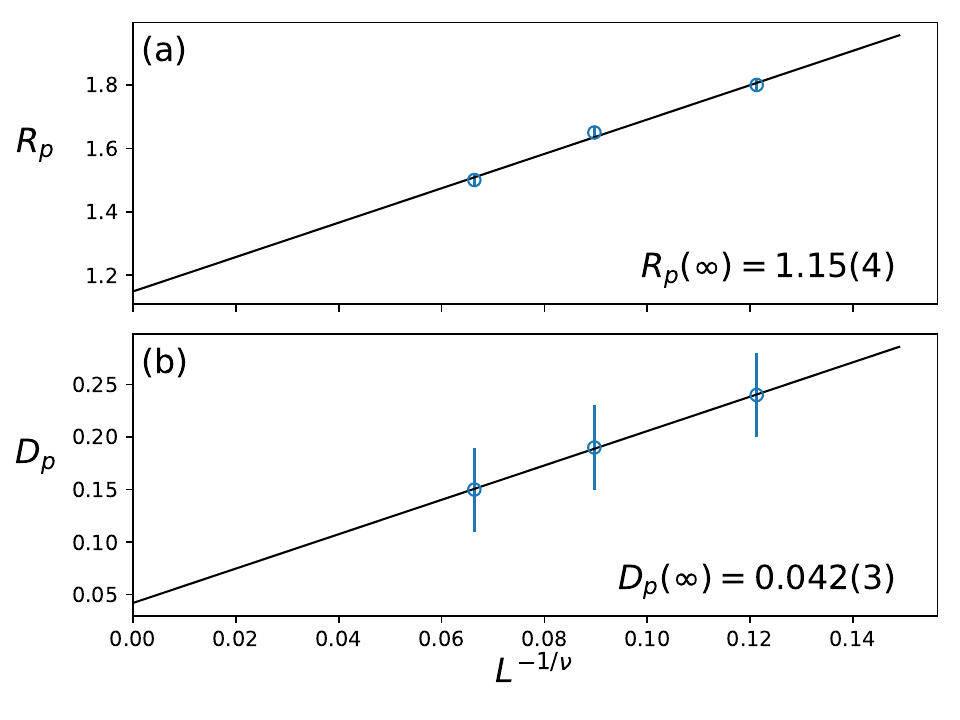}}
\caption{Finite-size scaling of the location $(R_p, D_p)$ at which the twist order parameter reaches its minimum
 at $\overline{z}_L=0$. The multicritical point in the thermodynamic limit is estimated as $R_p\approx 1.15$ and
$D_p\approx 0.04$ from an extrapolation to $L\to\infty$, using Eqs.~(\ref{eq:R_p}) and ~(\ref{eq:D_p}) with $\nu=2.3$.
}
\label{fig:P3}
\end{figure}

\begin{figure}[tbp!]
\centerline{\includegraphics[width=8.6cm]{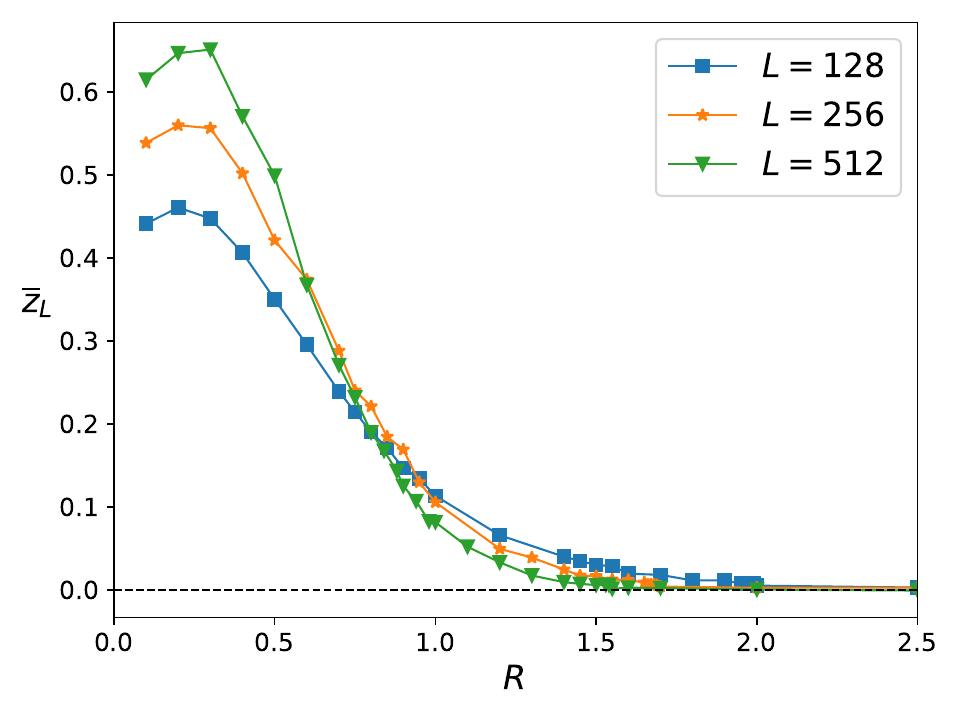}}
\caption{Disorder-averaged twist order parameters $\overline{z}_L$ for different system sizes
at $D=0$, plotted versus the randomness $R$.
}
\label{fig:z_D000}
\end{figure}

Here we calculate $z_L$ using the tSDRG method for the random spin-2 chain
with periodic boundary conditions.
We explore the behavior of the disorder-averaged
order parameter, $\overline{z}_L$, 
for a wide range of randomness and dimerization, parametrized by $R$ and $D$.
We have considered system sizes up to $L=512$ and more than 5,000 random coupling samples to obtain
the disorder average. 
Figure~\ref{fig:z_D} shows the disorder-averaged order parameter for different system sizes at $R=1.5, 1.0$, and $0.5$ 
as a function of $D$.
Here, in the cases with $R=0.5$ and $R=1.0$, one can clearly observe that the order parameter changes
its sign at certain values of $D$; there are two such zero-crossings for each $L$ that can be identified as 
two phase transition points  between different random VBS states.
The sign of $\overline{z}_L$ indicates that the domains from left (small $D$) to right (large $D$) 
are successively in $(2,2)$, $(3,1)$ and $(4,0)$-dominant phases.
By varying $R$, the two domain walls, located at the zero-crossings of $\overline{z}_L$, 
form critical lines that connect the two critical points in the clean limit, i.e.
$D_{c,1}\approx 0.18$ and $D_{c,2}\approx 0.55$  at $R=0$.  
The distance between these two critical lines decreases as $R$ increases.
For the case with $R=1.5$, one can observe the minimum value of $\overline{z}_L$ for $L=512$ (the largest system size that
we consider here) is about zero; we can than identify the associated $(R_p, D_p)$ with this minimum as the junction of
the two critical lines for the finite chain.
In Fig.~\ref{fig:z0}, the point at $(R_p, D_p)$ is denoted by $\tilde{P}_3^>$. 

The results described above for the disorder-averaged twist order parameter suggest that
there is a multicritical point, at which three phases $(2,2)$, $(3,1)$ and $(4,0)$ meet, 
in the region of $D\neq 0$. 
This implies that the diagram in Fig.~\ref{fig:diagram}(c) corresponds to the $R$-$D$ phase diagram of our model.
To determine the location of this multicritical point in the limit of $L\to\infty$,
we find the $(R_p(L), D_p(L))$ point  
at which the disorder-averaged twist order parameter reaches its minimum at $\overline{z}=0$,
and then estimate the critical values for $R_p$ and $D_p$ from an extrapolation to $L\to\infty$ using
\be
        R_p(L)-R_p(\infty) \sim L^{-1/\nu}\,,
        \label{eq:R_p}
\ee
and
\be
        D_p(L)-D_p(\infty) \sim L^{-1/\nu}\,,
        \label{eq:D_p}
\ee  
with $\nu=2.3$ for a multicritical point $\mathcal{P}_3$ (see Eq.~(\ref{eq:nu})).  
By doing so, we obtain $R_p(\infty)\approx 1.15$ and $D_p(\infty)\approx 0.04$, 
as shown in Fig.~\ref{fig:P3}.
{Here we round to two decimal places to point up $D_p(\infty)\neq 0$,
which is the chief characteristic of the phase diagram in Fig.~\ref{fig:diagram}(c).}
We note that, for a control parameter $\lambda$, the deviation of a finite-size pseudocritical point $\lambda_c(L)$
from the true critical point $\lambda_c(\infty)$ in the limit of $L\to\infty$ is often parameterized as $\ln(\lambda_c(L))-\ln(\lambda_c(\infty))$ for 
an infinite randomness fixed point~\cite{Fisher_AF,Fisher_Ising,XX,FFS}. 
For the bond strength distributions given in Eq.~(\ref{eq:J}) and Eq.~(\ref{eq:PJ}),
the distance from the critical point defined in Eq.~(\ref{eq:R_p}) is thus consistent with  
such logarithmic parametrization; the distance given in Eq.~(\ref{eq:D_p}) is also suitable
because the values of $D$ we consider here are so small that the approximation $\ln(1+D) \approx D$ is valid.

Now we turn to the undimerized region with $D=0$. Figure~\ref{fig:z_D000} shows
the dependence of disorder-averaged twist order parameters on $R$. Here the
twist order parameter for each system size $L$ is positive before converging to
$\overline{z}_L=0$ in the large $R$ region, consistent with the scenario in
which the system's ground state changes from a $(2,2)$ phase to an RS phase
when the randomness exceeds a critical value; this critical point at $D=0$ is also a
multicritical point $\mathcal{P}_3$ at which three phases $(2,2)$, $(4,0)$ and
$(0,4)$ meet  (see Fig,~\ref{fig:diagram}(c)).  Since the order parameter
$\overline{z}_L$ does not change its sign for the $(2,2)$-RS phase transition
nor for a nearby $(2,2)$-$(4,0)$ transition, it is difficult to determine the
transition point accurately by $\overline{z}_L$.  Nevertheless, the results
shown in Fig.~\ref{fig:z_D000} (and also in Fig.~\ref{fig:z_D} (a)) suggest
that the critical $R$ value at $D=0$ for the largest size $L=512$ is about
$R\approx 1.5$ (or slightly higher). 
Thus the multicritical point at $D=0$ (denoted as $\tilde{P}_3^0$ in Fig.~\ref{fig:z0}) and
the one at $D>0$ ($\tilde{P}_3^>$ in Fig.~\ref{fig:z0}) have critical $R$ values that are not far apart.


As seen in Fig.~\ref{fig:z_D}(c) and Fig.~\ref{fig:z_D000}, there are intersection points developing at some
nonzero values of $\overline{z}_L$.
In Ref.~\cite{Twist_spin1}, such intersection points of $\overline{z}_L$, instead of zero-crossings, were used to identify 
the multicritical point and the RS critical line for the random spin-1 chain.
Here, by exploring a wide range of parameters $(R, D)$ and accessing larger system sizes, we have found
that the $\overline{z}_L$ curve crossings appear only in the region of small $R$, or
the crossing points tend toward $\overline{z}_L=0$.   
Thus, the zero order parameter, $\overline{z}_L=0$, turns out to be a more reasonable indicator
for a transition point even in disordered systems.

\subsection{End-to-end correlations}
\label{sec:cor}

\begin{figure}[tbp!]
\centerline{\includegraphics[width=8.6cm]{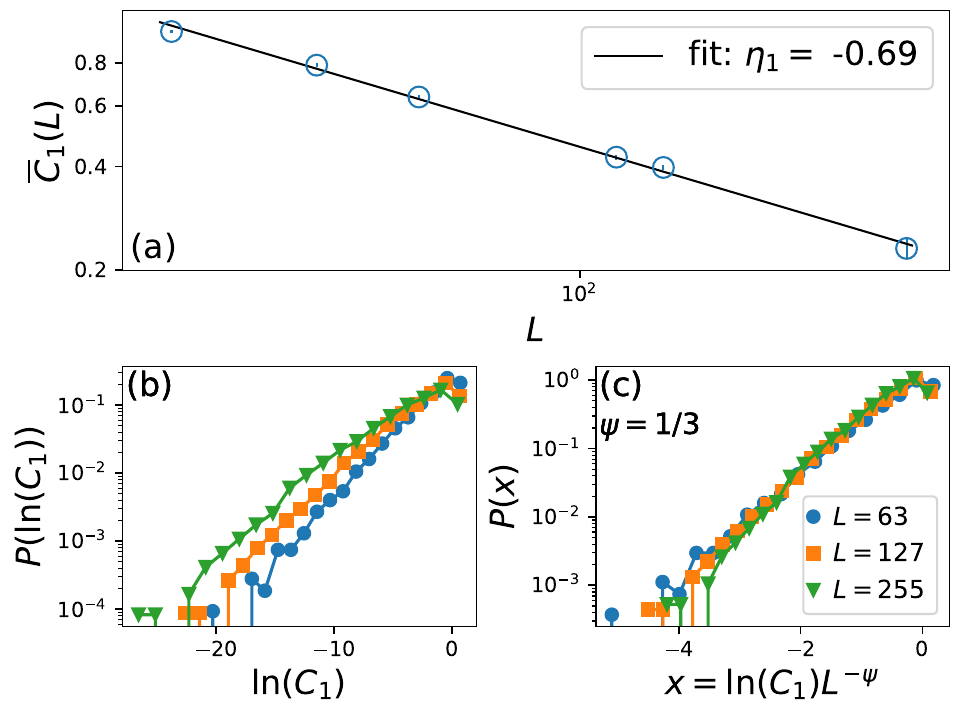}}
\caption{End-to-end correlations at the multicritical point $(D=0.04, R=1.15)$. (a): Finite-size dependence of the average correlation;
the solid red line has a slope of $\eta_1\approx 0.69$.
(b):  Distributions of end-to-end correlations for different sizes. 
(c): A scaling plot for the data in (b), assuming $-\ln(C_1) \sim L^\psi$ with $\psi=1/3$. 
}
\label{fig:c1_P3}
\end{figure}

In this subsection we investigate the end-to-end correlations in an open chain,
which considers correlations between two end spins.
This quantity is defined as
\be
    C_1(L)=(-1)^{L-1}\trip{\Psi_\text{GS}}{\mathbf{S}_1\cdot\mathbf{S}_L}{\Psi_\text{GS}}\,,
\ee
for the ground state $\ket{\Psi_\text{GS}}$ of a chain with $L$ spins and open boundary conditions.

\begin{figure}[tbp!]
\centerline{\includegraphics[width=8.6cm]{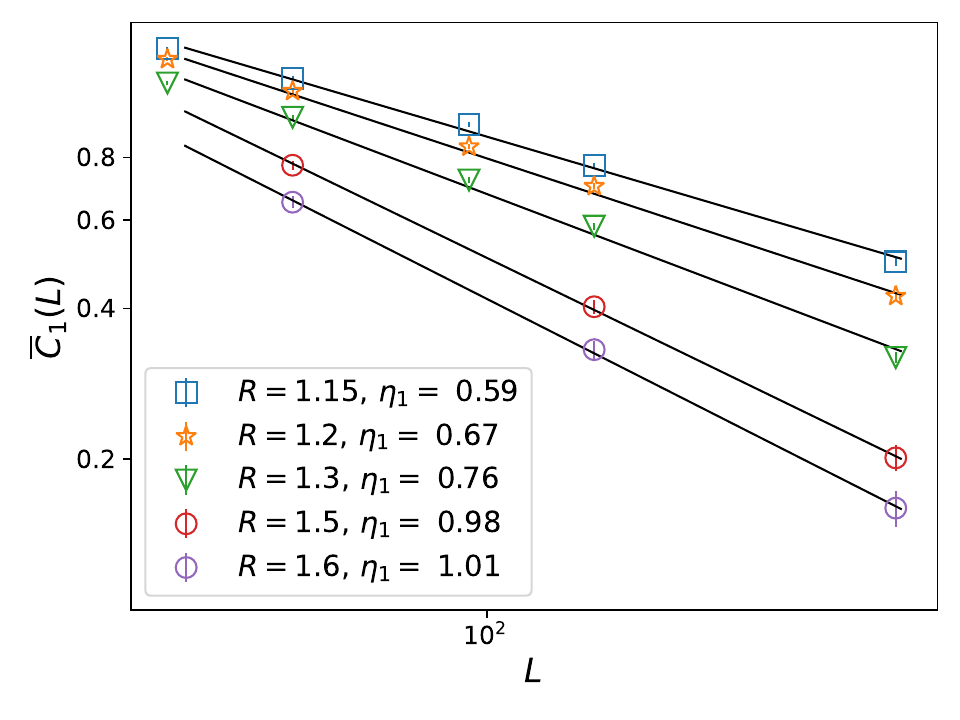}}
\caption{Average end-to-end correlation versus system size for the undimerized case ($D=0$) with various $R$ in a log-log plot.
All lines are fits to the form $aL^{-\eta}$. From the slopes of the fitting lines, corresponding to the exponents $\eta_1$,
we estimate $R\approx 1.2$ for the multicritical $\mathcal{P}_3$ point. For $R\geq 1.5$, the slope approaches $\eta_1=1$,
indicating that the system enters an RS phase; here it is the RS$_2$ phase.
}
\label{fig:c1_D0}
\end{figure}
   
Here we first summarize some previous results for the end-to-end correlations.
The end-to-end correlation of an open chain is closely related to the energy gap~\cite{Igloi_RW,Fisher_Young}.
In the RS phase the end-to-end correlations are very broadly distributed, with the typical behavior~\cite{Fisher_AF},
\be
   -\ln C_1(L) \sim L^\psi\,,
  \label{eq:Ctyp}
\ee
where $\psi=1/2$. On the other hand, the average end-to-end correlation function
decays as a power of $L$ at criticality~\cite{Fisher_Young},
\be
    \overline{C}_1(L) \sim 1/L^{\eta_1}\,,
    \label{eq:Cav}
\ee
with $\eta_1=1$ in the RS phase.
The behavior of the average end-to-end correlation in Eq.~(\ref{eq:Cav}) was first derived for
the infinite-randomness fixed point of the random transverse-field Ising spin chain~\cite{Fisher_Young}
and is also valid for the RS phase, as verified numerically in Ref.~\cite{Lajko,Tsai}.
At the multicritical point $\mathcal{P}_3$, typical end-to-end correlations go like Eq.~(\ref{eq:Ctyp})
but with $\psi=1/3$, according to SDRG analytical results. 
There have been so far no theoretical conjectures about the exponent in Eq.~(\ref{eq:Cav}) 
for the average end-to-end correlations at $\mathcal{P}_3$; nevertheless, previous numerical results~\cite{Lajko,Tsai}
for the spin-1 chain estimated $\eta_1 \approx 0.69-0.7$ for the $\mathcal{P}_3$ multicritical point,
which is expected also for the $\mathcal{P}_3$ point in the spin-2 chain.
Away from an infinite-randomness critical point, the distribution of end-to-end correlations has 
a power-law tail that behaves as~\cite{Igloi_RW}
\be
      P(C_1) \sim C_1^{-1+1/z}\,,
      \label{eq:c1_z}
\ee
with a finite dynamic exponent $z$; in a Griffiths phase, 
the singular low-energy behavior of various observables is characterized by a large and continuously variable dynamic exponent $z>1$.

\begin{figure}[tbp!]
\centerline{\includegraphics[width=8.6cm]{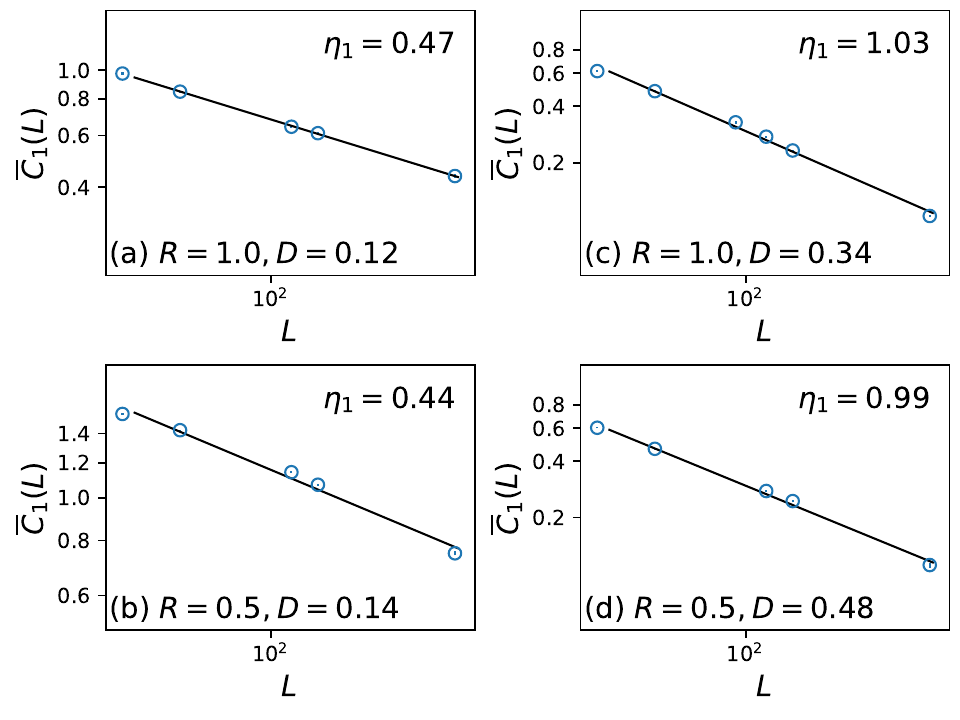}}
\caption{Finite-size dependence of the average end-to-end correlations at critical points, 
determined by the zero-crossings of the order parameter $\overline{z}_L$ for $L=512$,
in the region of $D>0$ at the (2,2)-(3,1) phase boundary [(a) and (b)] and the (3,1)-(4,0) boundary [(c) and (d)].
The mean correlations have a power-law decay; the exponent $\eta_1$ at the (3,1)-(4,0) boundary appears to approach $1$, 
consistent with the theoretical prediction for the RS$_{1/2}$ phase, while $\eta_1$ at the (2,2)-(3,1) boundary is much smaller than $1$.
}
\label{fig:c1_4D}
\end{figure}

We first examine the behavior of end-to-end correlations at the $\mathcal{P}_3$ points.
For the dimerized cases, we consider open chains with odd numbers of spins to balance the numbers
of strong and weak couplings.
Figure~\ref{fig:c1_P3}(a) shows the average of
the correlations at $R=1.15$ and $D=0.04$ (the point $P_3^>$ in Fig.~\ref{fig:z0}), where the location of the
multicritical point $\mathcal{P}_3$ is according to finite-size scaling of the
twist order parameters discussed in Sec.~\ref{sec:twist}. Here we estimate
$\eta_1\approx 0.69$ from the results for the average correlation as a function
of the chain length $L$, in good agreement with previous numerical results~\cite{Lajko,Tsai}.  
Also the distributions of the logarithmic correlations (shown in Fig.~\ref{fig:c1_P3}(b)), 
which become broader with increasing size, can collapse onto each other by using the scaled variable
\be
      x=\ln C_1/L^\psi,
      \label{eq:x}
\ee     
with $\psi=1/3$ [Fig.~\ref{fig:c1_P3}(c)], consistent with the theoretical prediction~\cite{Damle_Huse}.

For the undimerized case $D=0$, we show the average correlations plotted against $L$ for various values of $R$ in Fig.~\ref{fig:c1_D0};
all data here decay as a power law: $C_1(L)\sim L^{-\eta_1}$.
From the slopes of the fitting lines, corresponding to the exponents $\eta_1$,  
we estimate $R\approx 1.2$ for the multicritical $\mathcal{P}_3$ point (denoted by $P_3^0$ in Fig.~\ref{fig:z0}) by comparing the value of $\eta_1$ with previous numerical results
for the $\mathcal{P}_3$ point in the random spin-1 chain~\cite{Lajko,Tsai}. 
For stronger randomness, such as $R=1.5$ and $R=1.6$, the slopes approach $\eta_1=1$, 
which is the theoretical value for an RS phase;
in this case, it is the RS$_2$ phase.

\begin{figure}[tbp!]
\centerline{\includegraphics[width=8.6cm]{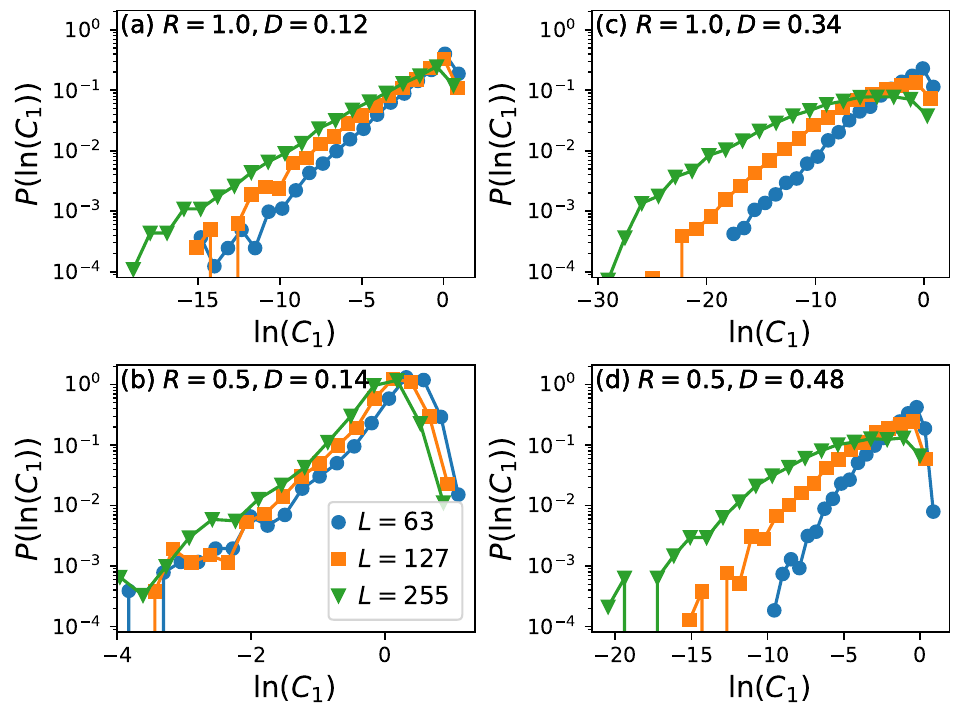}}
\caption{The distributions of the end-to-end correlations for different sizes at the same critical points as investigated in Fig.~\ref{fig:c1_4D}. 
The distributions for $(R=1.0, D=0.12)$ [Subfigure (a)] located at the (2,2)-(3,1) boundary with sufficiently strong disorder
and the distributions at the (3,1)-(4,0) boundary  [(c) and (d)] are broad and become broader with increasing $L$.
The correlations for $(R=0.5, D=0.14)$  [(b)] at the (2,2)-(3,1) boundary with smaller $R$, on the other hand, are not very broadly distributed,
and the data curves for different sizes are very similar, implying the critical phase is not of infinite-randomness type.
}
\label{fig:c1_4D_hist}
\end{figure}

\begin{figure}[tbp!]
\centerline{\includegraphics[width=8.6cm]{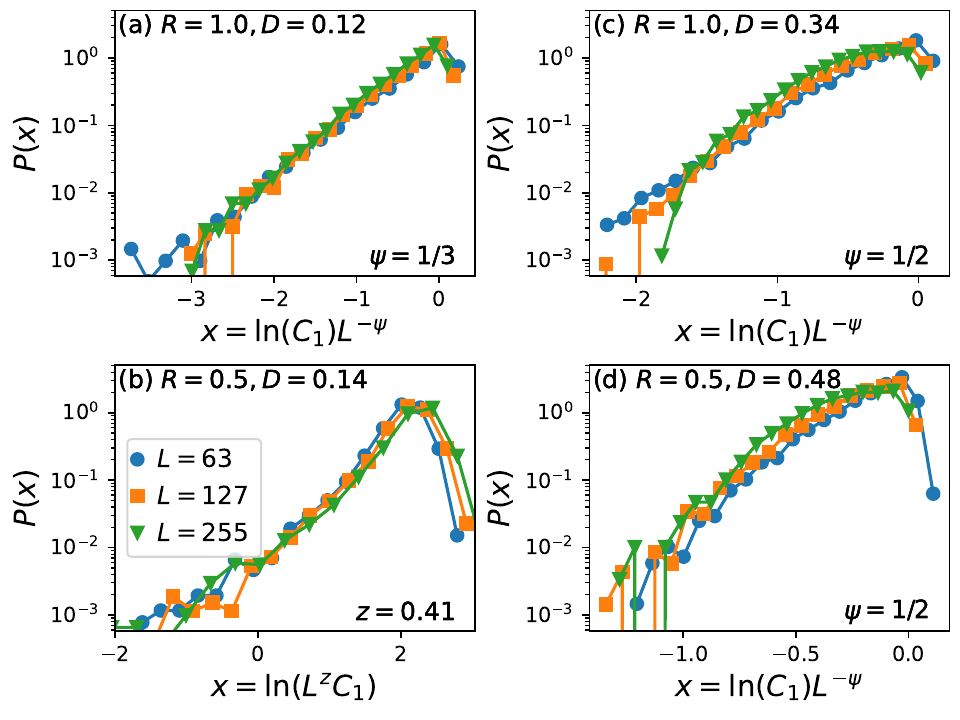}}
\caption{Scaling plots of the distributions in Fig.~\ref{fig:c1_4D_hist}.
The correlations in (c) and (d), at the the (3,1)-(4,0) phase boundary, are rescaled as $\ln C_1 /L^\psi$ with  $\psi=1/2$,
and the data in (a) for a critical point at the (2,2)-(3,1) phase boundary are plotted in terms of the same rescaled variable but with $\psi=1/3$ to
achieve good data collapse.
The correlations in (b), for a critical point at the (2,2)-(3,1) phase boundary with smaller $R$ and $D$, are
rescaled by assuming $P(C_1) \sim C_1^{-1+1/z}$  with $z=0.41$, which is estimated from the slopes of the tails
at small values of $C_1$. 
}
\label{fig:c1_4D_sc}
\end{figure}

Now we turn to the correlations on the critical lines at $D>0$ that are the (2,2)-(3,1) and (3,1)-(4,0) boundaries,
determined using the zero-crossings of the $\overline{z}_L$ order parameter for the largest system
size $L=512$. {Specifically, we focus on the four points denoted by $a,b,c$ and $d$ in Fig.~\ref{fig:z0}.}
Figure~\ref{fig:c1_4D} and Figure~\ref{fig:c1_4D_hist} show the averages and the distributions of end-to-end correlations 
at four such finite-size critical points. 
For the critical points, $(R=1.0, D=0.34)$ and $(R=0.5, D=0.48)$, at the boundary between the $(3,1)$ and $(4,0)$ phases,
our numerical results are found to be in good agreement with the analytic predictions for an RS phase: the average end-to-end correlations
decay as $\overline{C}_1(L)\sim 1/L$, as shown in Fig.~\ref{fig:c1_4D}(c) and (d), and the distributions of the correlations, 
shown in Fig.~\ref{fig:c1_4D_hist}(c) and (d), are extremely broad and become broader with increasing $L$.
Furthermore, the distributions for different system sizes collapse well onto each other by using the scaled variable defined in Eq.~(\ref{eq:x})
with the theoretical value $\psi=1/2$ for the RS phase, as shown in Fig.~\ref{fig:c1_4D_sc}(c)and (d). 

On the other hand, the results for the (2,2)-(3,1) phase boundary are not fully compatible with the
scenario of an RS phase.   
The mean end-to-end correlations $\overline{C}_1(L)$ at the (2,2)-(3,1) phase boundary, as shown 
in Fig.~\ref{fig:c1_4D}(a) and (b) for the two critical points $(R=1.0, D=0.12)$ and $(R=0.5, D=0.14)$,
appear to decay much slower than $1/L$, contrary to the behavior in an RS phase.
In particular, the distribution of the correlations in the region of small $R$, such as at $(R=0.5, D=0.14)$ (see Fig.~\ref{fig:c1_4D_hist}(b)),
does not broaden with increasing size, implying that the phase is not associated with an infinite-randomness critical point.
Assuming the scaling form given in Eq.~(\ref{eq:c1_z}), we estimate the dynamic exponent $z\approx 0.41$ from the slope of the power-law tail
of the distribution, which gives the scaling plot shown in Fig.~\ref{fig:c1_4D_sc}(b). 
With stronger disorder at $(R=1.0, D=0.12)$, the distribution becomes broader with increasing size (see Fig.~\ref{fig:c1_4D_hist}(a)),
showing the signature of infinite randomness.
We have used the scaled variable in Eq.~(\ref{eq:x}) and $\psi=1/3$
to achieve good data collapse, as shown in Fig.~\ref{fig:c1_4D_sc}(a).
Strong finite size effects and the close distance from the multicritical point at $R=1.15$ may lead to
the discrepancy between the exponent $\psi$ found here and the theoretical predicted value $\psi=1/2$ for an RS$_{1/2}$ phase. 

\begin{figure}[tbp!]
\centerline{\includegraphics[width=8.6cm]{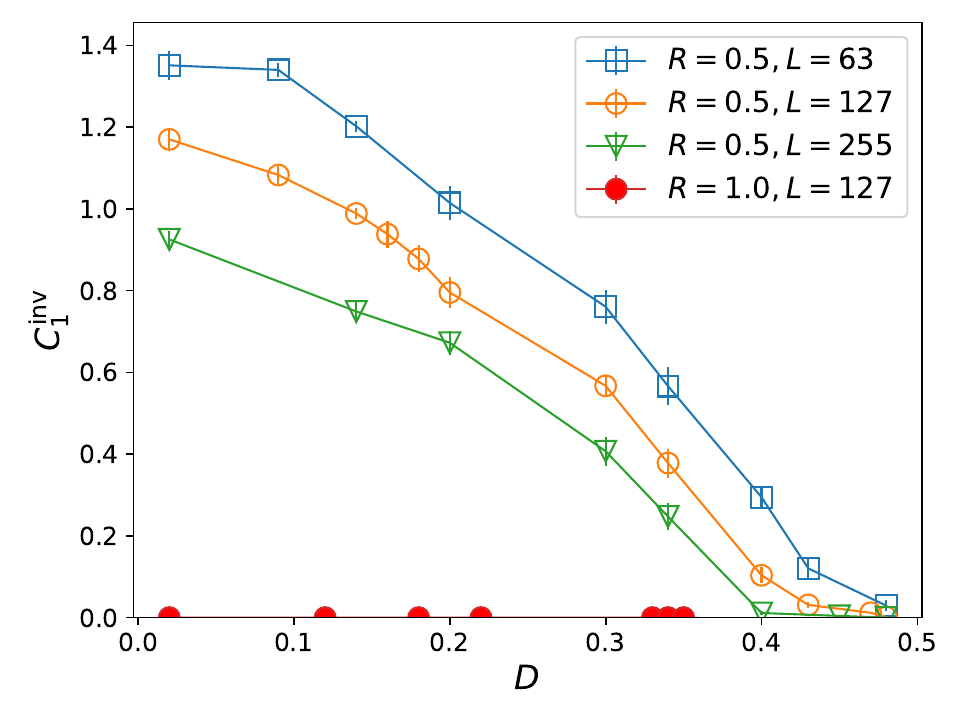}}
\caption{The inverse average of the end-to-end correlation, defined in Eq.~(\ref{eq:invC}), versus the strength of
dimerization $D$. 
}
\label{fig:invC}
\end{figure}

The small dynamic exponent $z<1$ estimated for $(R=0.5, D=0.14)$ 
does not lead to divergence of the local susceptibility~\cite{Vojta_rev,Lajko},
indicating there is no pronounced disorder-induced singular behavior.
Another way, suggested in  Ref.~\cite{Lajko}, to identify a nonsingular region with $z<1$ or a singular region with $z>1$ 
is via the so-called inverse average of the end-to-end correlation, defined as
\be
      C_1^{\text{inv}} = \left( \overline{C_1^{-1}} \right)^{-1}\,,
      \label{eq:invC}
\ee
where $\overline{C_1^{-1}}$ is the average of the inverses. 
The inverse average of $C_1$ is finite, $C_1^{\text{inv}}>0$ if $z<1$ , while 
it is zero if $z>1$. 
As shown in Fig.~\ref{fig:invC}, 
the inverse average $C_1^{\text{inv}}$ at $R=0.5$
is finite for a wide range of $D$ before it converges to zero at $D\approx 0.4$ for larger $L$,
while the inverse average vanishes at $R=1$ independent of $D$.
The results for the inverse average $C_1^{\text{inv}}$ also imply that there is no random-singlet phase at $R=0.5$
in the region with small dimerization $D<0.4$. 
This large nonsingular region also  gives rise to the weak signature of infinite randomness 
in the larger $R$ region at the (2,2)-(3,1) phase boundary.

\section{Summary and Discussion}
\label{sec:summary}

Using a tensor-network SDRG method, we have explored the ground-state phases of
the random spin-2 antiferromagnetic chain with alternating bond strength
distributions.  We have calculated the twist order parameter, defined as the
ground-state expectation value of the unitary operator in Eq.~(\ref{eq:U}), to
classify the types of random VBS phases depending on the strength of bond
randomness $R$ and the dimerization $D$.  For a disorder-free VBS phase
$(\sigma,4-\sigma)$ in a clean system, the twist order parameter is positive if
$\sigma$ is even and negative if $\sigma$ is odd~\cite{Twist}.  In a random VBS
domain, there is nonzero residual VBS order (dimerization) that can be detected
by the disorder average of this order parameter, as we have demonstrated in
this paper.  Therefore, the zero-crossing of the disorder-average twist order
parameter can serve to determine the phase transition point between different
random VBS states, in the same way as for clean systems~\cite{Twist}.  Our
results largely agree with the phase diagram sketched in
Fig.~\ref{fig:diagram}(c).  There is a multicritical point in the intermediate
disorder regime with finite dimerization, where (2,2), (3,1) and (4,0) three
phases meet.  The (2,2)-(3,1) phase boundary and the (3,1)-(4,0) boundary
extend to $R=0$ and are predicted to be in the RS$_{1/2}$ phase for any
$R>0$~\cite{Damle_Huse}.  However, from the results for end-to-end correlations
we see no signs of an RS phase at the (2,2)-(3,1) boundary with small $R$ and
have instead found a large nonsingular region, characterized by $z<1$, in the
small $R$ regime.  Such a nonsingular region with $z<1$ in the weak disorder
limit has previously also been identified in numerical studies of the random
undimerized ($D=0$) spin-1 chain~\cite{Lajko,Tsai} and spin-3/2
chain~\cite{S32_Carlon}; these studies used the same power-law distribution of
bond randomness, which is also identical to the distribution we consider here
for the undimerized case.

\begin{figure}[tb!]
\centerline{\includegraphics[width=8.6cm]{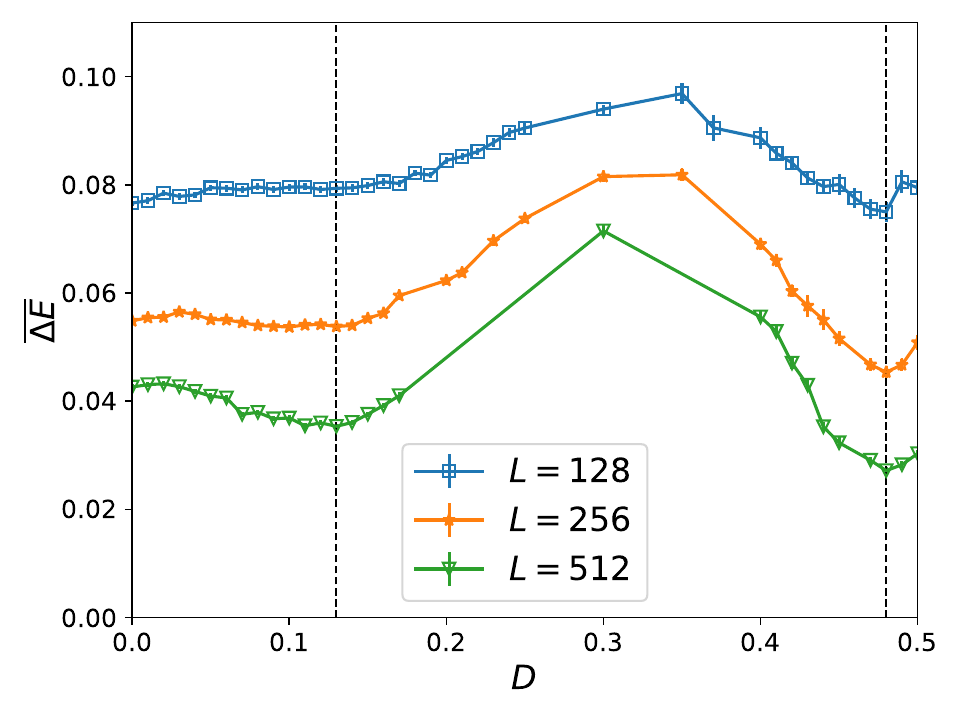}}
\caption{The average gap at $R=0.5$ as a function of the strength of
dimerization $D$. The gap for each sample is obtained from the lowest-lying energy gap of the top Hamiltonian $H_\text{top}$
in the tree tensor network. The two dashed lines indicate
the $(2,2)$-$(3,1)$ boundary (left line) and the $(3,1)$-$(4,0)$ boundary (right line),
based on the zero crossings of $\overline{z}_L$ for $L=512$.
}
\label{fig:gap}
\end{figure}

The nonsingular behavior in the weak disorder and weak dimerization limit 
and the absence of the RS$_{1/2}$ phase there can have more than one source.
Compared to the Haldane gap value of $\approx 0.41J$ in the spin-1 case with nearest-neighbor coupling $J$,
the finite gap in the clean spin-2 chain is much smaller, just about $\approx 0.09J$.  
Even so, this small energy gap appears to have considerable impact on the ground-state properties of the (2,2) phase at weak disorder.
Furthermore, 
we did not set a constraint to the strength difference between odd and even bonds while using the bond strength distribution 
(given in Eq.(\ref{eq:J})) in our calculations;
this may produce unsharp dimerization for small systems, especially in the small $D$ limit,
which in turn leads to a weak critical signature or the absence of the RS$_{1/2}$ phase at the (2,2)-(3,1) boundary.    
In this regard we present in Fig.~\ref{fig:gap} the average energy gap of the chain with PBC at $R=0.5$, 
obtained from the lowest-lying excitation of the top Hamiltonian $H_\text{top}$,  
versus the dimerization $D$.
Here we observe a clear (local) minimum in the average energy gap at $D = 0.48$ for all system sizes, with the $(3,1)$-$(4,0)$ transition point indicated by the right dashed line --- estimated from the zero crossing of the twist
order parameter.
On the other hand, 
in the small $D$ region (corresponding to the nonsingular region), 
the curve for the largest size $L=512$ develops a less clear minimum around the estimated (2,2)-(3,1) transition point (indicated by the left dashed line),
while the curves for smaller sizes are flat in this region, showing strong finite-size effects.

The twist order parameter 
is a useful quantity, also for disordered systems, to determine the phase transition point between different random VBS states
based on changes of the sign according to the valence-bond configuration.
However, the phase diagram of the spin-2 chain considered here has a (2,2)-(4,0) phase boundary where
the twist order parameter does not change the sign, which makes it difficult to detect this phase transition.
It would be desirable to find an order parameter that can accurately determine the (2,2)-(4,0) phase boundary 
which connects two $\mathcal{P}_3$ multicritical points.

Recently, there has been an increasing interest in properties of higher-spin
materials both from the theoretical and experimental perspective, especially in
the context of Kitaev models and quantum spin
liquids~\cite{Kitaev_Rev,Trebst,Kee_PRL}.  Like the spin-2 chain that we
consider here, the quasi-one-dimensional versions of these higher-spin
materials exhibit rich phase diagrams~\cite{Sen,Kee_1D}.
Concerning disorder effects on the low-temperature phases, the tSDRG method and
its variants~\cite{Tree2017,Tree2020_PRB,Tree2021_PRB,Tree2022_PRB,Tree2023_PRR,Tree2023}
are certainly the most promising numerical tools for studying large-scale systems with accuracy.

\begin{acknowledgments}
{We would like to thank F.~Igl\'oi and  J.~Hoyos for useful discussions.} 
This work was supported by the National Science and Technology Council (NSTC) of
Taiwan under Grants No. 111-2119-M-007-009, 112-2119-M-007-008,
112-2112-M-004-008, 111-2112-M-004-005, 110-2112-M-007-037-MY3.  We also
acknowledge support from the NCTS.
\end{acknowledgments}

\end{document}